\documentclass[prl,a4paper,superscriptaddress,twocolumn,showpacs,amsmath,amssymb,floatfix,amstex]{revtex4}

\usepackage[dvips]{graphicx}
\usepackage{bm,pstricks,longtable}

\input{epsf}

\begin{document}

\title{Destruction of  Anderson localization by a weak nonlinearity}
\author{A.S.Pikovsky} 
\affiliation{\mbox{Department of Physics, University of Potsdam, 
  Am Neuen Palais 10, D-14469, Potsdam, Germany}}
\author{D.L.Shepelyansky}
\affiliation{\mbox{Laboratoire de Physique Th\'eorique, UMR 5152 du CNRS, 
  Universit\'e  Toulouse III, 31062 Toulouse, France}}
\affiliation{\mbox{Department of Physics, University of Potsdam, 
  Am Neuen Palais 10,  D-14469, Potsdam, Germany}}

\date{\today}

\begin{abstract}
We study numerically a  spreading of an initially localized
wave packet in a one-dimensional
discrete nonlinear Schr\"odinger lattice with disorder. 
We demonstrate that
above a certain critical strength of nonlinearity  
the Anderson localization is destroyed and an 
unlimited subdiffusive spreading of the field along the lattice occurs.
The second moment grows with time $ \propto t^\alpha$, 
with the exponent $\alpha$ being in the range $0.3 - 0.4$.
For small nonlinearities the distribution remains localized 
in a way similar to the linear case. 
\end{abstract}


\maketitle

The Anderson localization \cite{anderson1958} has been
originally discussed for electron propagation in a disordered
potential. Nowadays, an impressive technological progress in experiments 
with cold atoms
allows one to create a disordered quasi-1D 
potential by laser fields and to detect 
signatures of localization of a Bose-Einstein condensate (BEC) 
in presence of disorder
\cite{inguscio2005,aspect2005,inguscio12005,ertmer2005,aspect2006}.
An interesting new aspect 
in such systems is an importance of nonlinear effects since 
in a good approximation the evolution of BEC can be described
by the nonlinear Gross-Pitaevskii equation (see e.g. \cite{pitaevskii}).
An interplay of disorder, localization, and nonlinearity
appears also in other physical systems like
wave propagation in nonlinear disordered media
(see e.g. \cite{skipetrov2000,segev2007}), 
chains of nonlinear oscillators (see e.g. \cite{james2005})
with randomly distributed frequencies,  
and models of quantum chaos with a kicked soliton \cite{pikovsky1991}
and a kicked BEC \cite{phillips2006,summy2006,sadgrove2007}.

We focus here on the 
discrete Anderson nonlinear Schr\"odinger equation (DANSE)
\begin{equation}
i \hbar{{\partial {\psi}_{n}} \over {\partial {t}}}
=E_{n}{\psi}_{n}
+{\beta}{\mid{\psi_{n}}\mid}^2 \psi_{n}
 +V ({\psi_{n+1}}+ {\psi_{n-1})}\;,
\label{eq1}
\end{equation}
where 
$\beta$ characterizes nonlinearity,
$V$ is hopping matrix element, on-site
energies are randomly distributed in the range
$-W/2 < E_n < W/2$, and the total probability is normalized to unity
$\sum_n {\mid{\psi_{n}}\mid}^2 =1$. For $\beta=0$ and weak disorder
all eigenstates are exponentially localized
with the localization length $l \approx 96 (V/W)^2 $
at the center of the energy band \cite{kramer1993}. 
Hereafter we set for convenience $\hbar=V=1$, 
thus the energy coincides with the frequency. 

For nonlinear equation (\ref{eq1}) we
consider the following problem: how an initially localized field 
$\mid{\psi_{n}(0)}\mid^2=\delta_{n,0}$ is spreading?
In the linear case the spreading saturates after excitation of 
all linear modes that have significant values at $n=0$. 
The same process of ``initial excitation'' of modes happens in the nonlinear case as well, this
 initial stage of spreading has been analyzed recently in refs.~\cite{aspect2007,shapiro2007} and is now well
understood. However, a behavior at large time scales
remains less clear. The results presented in \cite{aspect2007}
support the view of eventual exponential localization of the field.
We demonstrate below that the spreading is unlimited, however it is rather slow -- subdiffusive. 

The basic idea is to use the equivalence between the Anderson localization 
and the localization of quasienergy eigenstates in 
a kicked quantum rotator~\cite{Fishman82,Shepelyansky-87}. In the latter model 
the case of quantum chaos 
with nonlinearity has been considered
analytically and numerically in \cite{dls1993,kottos2004} and it has been
shown that above a certain nonlinearity level, nonlinear phase shifts
lead to a complete delocalization with a subdiffusive spreading over
all states \cite{dls1993}.  Furthermore it has been argued that the same situation should 
appear for the DANSE (\ref{eq1}).

We first apply the theoretical arguments of paper \cite{dls1993} to model (\ref{eq1}), 
and then perform large scale numerical simulations
of a wave packet spreading on a time scale which is by 5-6 orders of magnitude
larger compared to that in \cite{aspect2007}. Our results
are in general consistent with the theory developed for the
quantum chaos model \cite{dls1993} that predicts unlimited subdiffusive spreading.

At first glance, the effect of nonlinearity seems 
to be vanishing in the limit of 
a broadly spread distribution. Indeed, if 
the field is spread over $\Delta n$ sites, 
then due to the conservation of the total probability in Eq.~(\ref{eq1}) 
the field is small $ \mid{\psi_{n}}\mid^2\approx 1/\Delta n$ and 
correspondingly small are the nonlinear effects. However, one should compare 
the nonlinear frequency shift $\delta \omega\sim \beta \mid{\psi_{n}}\mid^2$ 
with the characteristic distance $\Delta \omega$ between frequencies 
of excited modes (the latter are the
exponentially localized 
modes of the linear disordered lattice). 
As the number of these modes is proportional to $\Delta n$, the distance between the
frequencies obeys $\Delta \omega\sim1/\Delta n$, 
and the relative nonlinear frequency shift
$\delta \omega/\Delta \omega\sim \beta$ is 
independent of the width of the field distribution but is proportional 
to the nonlinearity parameter $\beta$. This means that the effect of nonlinearity 
does not qualitatively depend on the width of field distribution: 
if for some $\beta>\beta_c$ 
the field is chaotic, chaos remains while spreading, 
and no transition to regularity that blocks spreading is expected; 
for small nonlinearities $\beta<\beta_c$ the dynamics which is nearly regular 
(KAM regime) 
for localized distributions remains as such for all times. 
However, quantitative difference can occur and the spreading can slow down 
for wide distributions. Again, to roughly estimate this effect, 
we can adopt here the arguments of \cite{dls1993}. 
In the basis of linear localized modes, 
the evolution of the amplitudes $C_m$ of 
these modes is due to their nonlinear coupling, 
i.e. $\dot C_m\sim \beta C_{m_1} C_{m_2} C_{m_3}$. Assuming randomness of the phases, 
we can estimate the rate of excitation of a newly involved mode
as $\sim  \mid{C}\mid^6\sim 1/ (\Delta n)^3$. On the other hand, 
excitation of a new mode is none other than 
diffusive spreading of the field, thus
$\frac{d}{dt} (\Delta n)^2\sim 1/ (\Delta n)^3$. 
Solution of this equation yields subdiffusive spreading
\begin{equation}
(\Delta n)^2\propto t^{2/5}\;.
\label{eq10}
\end{equation}

For numerical simulations we used the operator splitting
integration scheme for the time evolution given by (\ref{eq1}):
$\psi_{n}(t+\Delta t) = \hat{R}\hat{S}\psi_{n} (t)$, where 
$\hat{R} = 
\exp(-i (E_n + \beta {\mid{\psi_{n}}\mid}^2) \Delta t)$ is local and 
$\hat{S}=\exp( - 2 \Delta t \cos {\hat {\theta}} ) $ 
is nontrivial because
$\hat {\theta}$ is the operator conjugated to 
${\hat n} = -i \partial / \partial \theta$. 
This kick-like integration scheme is unitary and 
preserves the total probability. In addition it introduces
high harmonics with frequencies $\omega_m = m 2\pi/\Delta t$
and integer $m$. However, at small $\Delta t$
these frequencies are significantly larger than
the system energy band width  $B$: 
$\omega_1 = 2\pi /\Delta t \gg B=(4+W)$
and their average effect should be exponentially small.
We have chosen $\Delta t = 0.1$ that gives high frequency
oscillations of total energy on a percent level.
A further decrease of $\Delta t$ by an order of magnitude
does not affect the average behavior of the field spreading.

We used two discrete implementations for the evolution operator $\hat{S}$
of the linear Schr\"odinger equation. In the first one we represent $\hat{S}$ 
as a band matrix whose elements are
Bessel functions.
At $\Delta t =0.1$ we keep Bessel functions $J_m(2 \Delta t)$ 
with $\mid m \mid \leq 10 $ that
preserves probability on one integration step
with the accuracy better than $10^{-16}$, whereas
after time $t=10^{8}$ the probability is preserved
with accuracy better than $10^{-7}$. For one disorder
realization such a run with the total number of sites $N=2001$
and $t=10^8$ 
takes about 6000 mins of CPU on a $4 GHz$-workstation.
In another implementation we used the unitary Crank-Nicholson scheme~\cite{numrec}.
The results obtained by both methods are very similar; 
below mainly those from the first method are presented because of its better efficiency.

\begin{figure}[h]
   \centering
   \includegraphics[width=0.45\textwidth,angle=0]{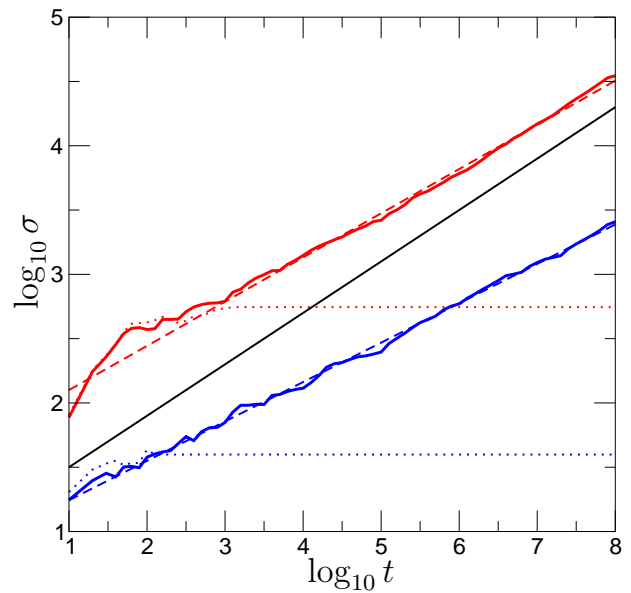}
\caption{(color online)
The dependence of the second moment $\sigma $ of the
probability distribution
$w_n$ on time $t$ for two disorder strengths
$W=2$ (top full red/gray curve) and 
$W=4$ (bottom full black/blue curve)
for $\beta=1$; dotted curves of same color
show data at $\beta=0$ (for large $t$ 
only the average value is shown by a horizontal line).
The values of $\log_{10} \sigma$ are averaged over 8 disorder realizations
and over  time intervals
$\Delta (\log_{10} t) \approx 0.1$.  
Dashed lines show the numerical fits
$\log_{10} \sigma = \alpha \log_{10} t + \eta$ 
with $\alpha=0.344 \pm 0.003$, $\eta=1.76 \pm 0.02$
(done for $3 \leq \log_{10} t \leq 8$ for $W=2$)
and
$\alpha=0.306 \pm 0.002$, $\eta=0.94 \pm 0.01$
(done for $2 \leq \log_{10} t \leq 8$ for $W=4$).
The full straight line shows the slope $\alpha=0.4$.
Here and below the logarithms are decimal.
%
\label{fig1}
}
\end{figure}

To characterize the wave packet spreading over the lattice sites
we compute its average squared width, i.e. the second moment
$\langle(\Delta n)^2\rangle=
\sigma(t) = \sum_n (n-\langle n\rangle)^2 {\mid \psi_{n}(t) \mid}^2$.
The averaging over disorder realizations was performed 
for the logarithm of this quantity, i.e. for
$\log \sigma$. The dependence of the averaged $\sigma$ on time $t$
is shown in Fig.~\ref{fig1} for a moderate nonlinearity $\beta=1$
and disorder strengths $W=2$ and $W=4$. It clearly shows a subdiffusive
spreading 
\begin{equation}
\sigma(t) \propto t^{\alpha}
\label{eq2}
\end{equation}
which continues without any sign of saturation 
up to extremely large time $t_{max}=10^8$.
At $t_{max}$  the variance $\sigma$ becomes by two orders of magnitude larger
than its saturation value at $\beta=0$. Yet the initial
spreading at $t \lesssim l$ for $\beta=1$ is similar to
the linear case $\beta=0$ in agreement with \cite{shapiro2007}. 
The exponent $\alpha$ was determined by a fit
in the time interval $t_0 < t <t_{max}$,
where $t_0$ is the characteristic time at which the linear spreading ends:
$\sigma(\beta=0;t) \lesssim \sigma(\beta,t_0)$.
The fits are shown in Fig.~\ref{fig1}
and the fit values are given in the caption to Fig.~\ref{fig1}.
The statistical error in the value of $\alpha$
is rather small due to a large time interval, 
however $\alpha$ values for individual realizations fluctuate
rather strongly varying in the interval $0.32 \leq \alpha \leq 0.39$
and $0.28 \leq \alpha \leq 0.41$ 
with the standard deviation errors $\Delta \alpha/\alpha = 0.026$ and $0.045$,
for $W=2$ and $4$ respectively. There are also certain
time variations, e.g. for $10^5 \leq t \leq 10^8$
the fits give $\alpha = 0.375$ and $0.319$ for $W=2$ and $4$ respectively.
In spite of these variations the value of $\alpha$
differs visibly from the theoretically expected 
value $\alpha=0.4$ (Ref.~\cite{dls1993} and Eq.~(\ref{eq10})).
We will return to the discussion of this deviation later.

\begin{figure}[h]
   \centering
   \includegraphics[width=0.45\textwidth,angle=0]{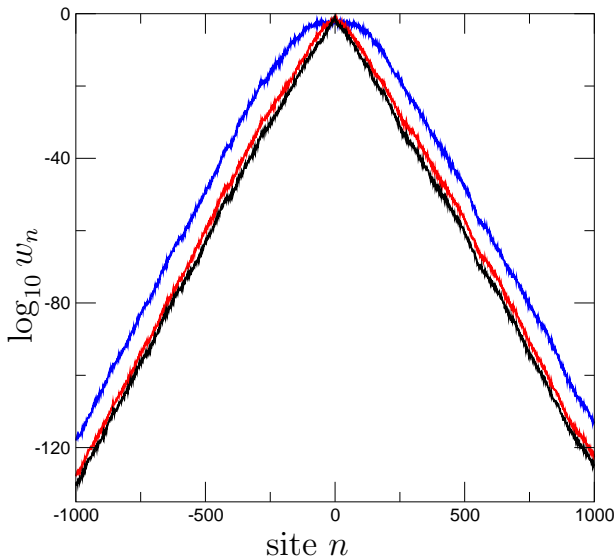}
\caption{(color online) Probability distribution $w_n$
over lattice sites $n$ at $W=4$ for $\beta=1$, $t=10^8$
(top blue/solid curve) and $t=10^5$  (middle red/gray curve);
$\beta=0, t=10^5$ (bottom black curve;
the order of the curves is given at $n=500$).
At $\beta=0$ a fit $\ln w_n = -(\gamma |n| + \chi$) gives
$\gamma \approx 0.3,\; \chi\approx 4$.
The values of $\log_{10} w_n$ are averaged over the same disorder realizations
as in Fig.~\ref{fig1}.
\label{fig2}
}
\end{figure}

\begin{figure}[h]
   \centering
   \includegraphics[width=0.45\textwidth,angle=0]{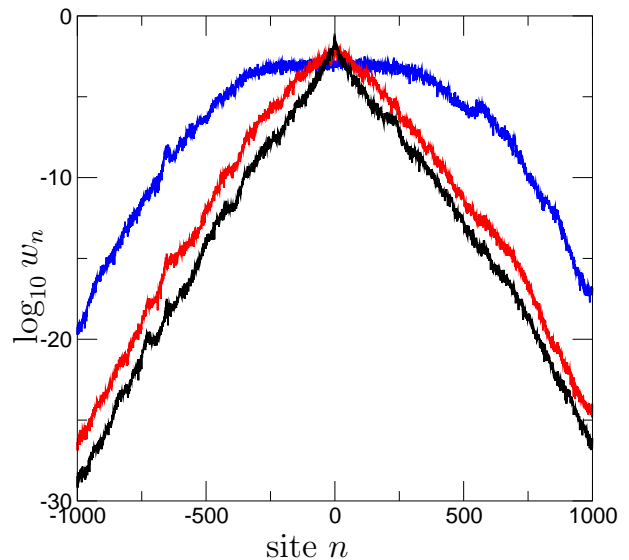}
\caption{(color online) 
Same as in Fig.~\ref{fig2} but with $W=2$.
At $\beta=0$ a fit $\ln w_n = -(\gamma |n| + \chi)$ gives
$\gamma \approx 0.06,\; \chi\approx -3$.
The values of $\ln w_n$ are averaged over the same disorder realizations
as in Fig.~\ref{fig1}.
\label{fig3}
}
\end{figure}

A more detailed characterization of the field spreading
is given by the probability distribution $w_n = {\mid {\psi_n (t)} \mid}^2$.
We show $\log_{10} w_n$ averaged over disorder realizations in 
Figs.~\ref{fig2},\,\ref{fig3} at $t=10^5$ and $10^8$.
For $W=4$ the tails of the probability distribution
drop down to enormously small values $\sim 10^{-130}$
that can be reached due to our integration scheme
which works efficiently up to very small absolute values of probability.
The tails of the distribution $w_n$ drop exponentially with the same
slope as for the linear case $\beta=0$ which is also shown;
the decay rate $\gamma \approx 0.30$ is close to the 
theoretical value $2/l \approx 0.33$.
Another notable feature of $w_n$ is a flat distribution, {\it chapeau},
centered near the initially populated site $n=0$. Inside the chapeau
the sites are populated in an approximately homogeneous way,
and hence its width is essentially determined by the second moment $\sigma(t)$.
For $W=2$ the decay rate $\gamma$ for $\beta=0$ 
drops approximately by a factor $\approx 4$ compared to the case $W=4$, 
in agreement with 
the theoretical expression for $l$. Due to a larger value of $l$,
the spreading over the lattice sites is broader and  a non-exponential
shape of the distribution $w_n$ at $t=10^8$ is more visible.
At shorter times $t=10^5$ the tails of the distribution are very similar
to those in the linear case $\beta=0$.

\begin{figure}[h]
   \centering
   \includegraphics[width=0.45\textwidth,angle=0]{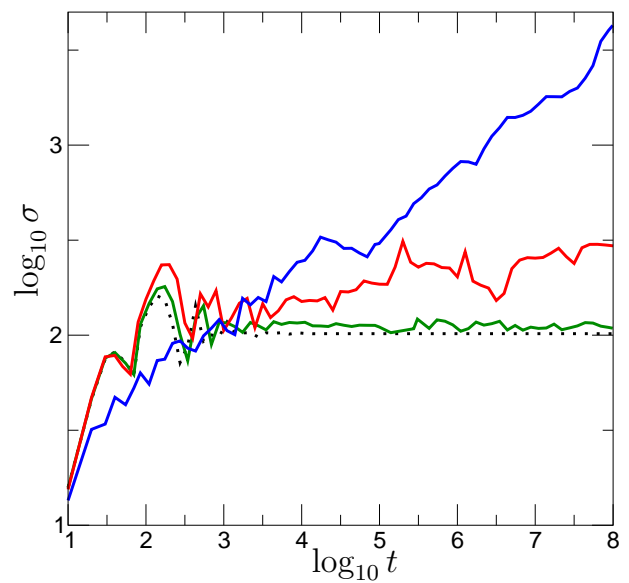}
\caption{(color online) 
Dependence of the second moment $\sigma$ of probability distribution
$w_n$ on time $t$ for different strengths of nonlinearity 
$\beta =1, 0.1, 0.03, 0$ at $W=4$
(curves from top to bottom
at $\log_{10} t =4.5$, respectively).
Data are shown for one particular disorder realization.
\label{fig4}
}
\end{figure}

The dependence of $\sigma(t)$ on nonlinearity $\beta$ 
is shown in Fig.~\ref{fig4} for $W=4$. One can clearly see that 
for $\beta=0.03$ the growth of $\sigma$ is stopped completely,
the probability distribution $w_n$ is close to exponential
localization like for $\beta=0$ (see Fig.5). For $\beta=0.1$
there is still a very slow increase of $\sigma$ with time,
which is however hardly distinguishable from a saturation.
This value of $\beta$ is presumably close to a critical one, 
at which the unlimited spreading sets on. 
A similar transition occurs for $W=2$.
These data show that
a delocalization transition takes place 
at a certain critical nonlinearity $\beta_c \approx 0.1$.
We note that the qualitative and quantitative features of the dynamics seem to be independent of the sign of $\beta$: the spreading for $\beta=-1$ is similar to that for $\beta=1$ (we have checked this also for $\beta=\pm 2$).

\begin{figure}[h]
   \centering
   \includegraphics[width=0.45\textwidth,angle=0]{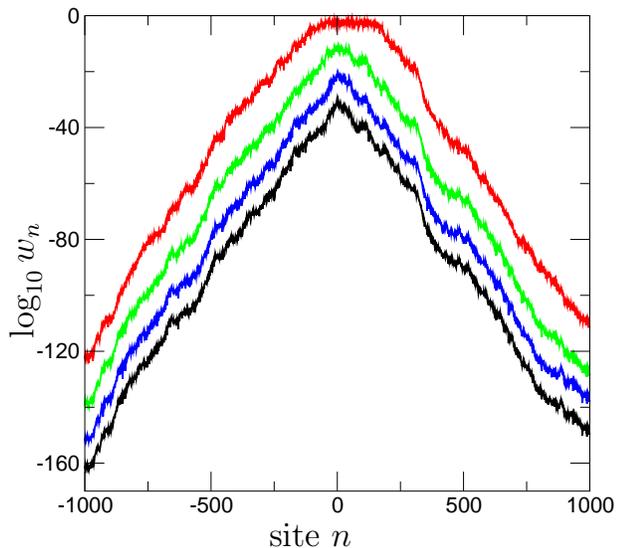}
\caption{(color online) Probability distribution $w_n$
over lattice sites $n$ for the same disorder realization as in Fig.~\ref{fig4},
for $W=4$, $t=10^8$ and for different nonlinearities
$\beta=1, 0.1, 0.03, 0$ (from top to bottom; for clarity
3 bottom curves have additional vertical shifts -10, -20 and -30
compared to the top one).
\label{fig5}
}
\end{figure}

The obtained results show close similarities between the DANSE model 
and the nonlinear kicked rotator studied in \cite{dls1993}.
In both cases for $\beta > \beta_c$ a subdiffusive spreading 
over the lattice continues up to enormously long times.
In both models the probability distribution  has a chapeau
with approximately homogeneous probability distribution.
Outside of the chapeau the probability drops exponentially.
The width of the chapeau grows in a subdiffusive way 
and the exponent $\alpha$ of this growth is approximately the 
same for both models. 
The question about the exact value of the exponent remains open.
It is possible that at $W=4$ the localization length is relatively short
and there are deviations from the theoretical value $\alpha=0.4$
given in \cite{dls1993}. 
Longer computations with
a better statistical averaging are needed to determine the
exact value of $\alpha$; the latter may also depend on the parameters $\beta$ and $W$.
In particular, slower diffusion might be due to inhomogeneities 
of the effective mode-to-mode hopping rates, which are more pronounced 
for smaller localization lengths, i.e. for larger disorder $W$.
At the same time the obtained numerical data clearly show the existence of
unlimited spreading over the lattice for $\beta > \beta_c$.
Indeed, for the data of Fig.~\ref{fig1} at $W=2$
the nonlinear frequency shift $\delta \omega \approx \beta w_n 
\approx \beta /\sqrt{\sigma(t_{max})} \approx 0.006 $ is much smaller than the
typical level spacing between localized states $\delta \nu \approx B/l \approx 0.25$.
The same is true for $W=4$. This means that in simulations we have 
reached the time scales
with apparently asymptotic behavior. 

In conclusion, we have demonstrated that in a one-dimensional
nonlinear disordered lattice the Anderson localization is destroyed and 
the field spreads subdiffusively far beyond the linear localization range. 
This effect appears to have a threshold in the nonlinearity coefficient, 
although the transition might be not perfect, 
as even for small nonlinearities an
extremely slow spreading of the field due 
to Arnold diffusion mechanism is not excluded. 
It appears also promising to look for other manifestations of nonlinearity-induced 
destruction of localization, e.g. in the scattering problem~\cite{tobepub}.

We thank S.Aubry and S. Flach for useful discussions.



\begin{thebibliography}{99}
\bibitem{anderson1958} P.W.~Anderson, Phys. Rev. {\bf 109}, 1492 (1958).
\bibitem{inguscio2005} J.E.~Lye, L.~Fallani, M.~Modugno, D.S.~Wiersma, 
        C.~Fort, and M.~Inguscio, Phys. Rev. Lett. {\bf 95}, 070401 (2005).
\bibitem{aspect2005} D.~Cl\'ement, A.F.~Var\'on, M.~Hugbart, J.A.~Retter, P.~Bouyer,
        L.~Sanchez-Palencia, D.M.Gangardt, G.V.Shlyapnikov, and A.~Aspect,
        Phys. Rev. Lett. {\bf 95}, 170409 (2005).
\bibitem{inguscio12005} C.~Fort, L.~Fallani, V.~Guarrera, J.E.~Lye, M.~Modugno,
        D.S.~Wiersma, and M.~Inguscio, Phys. Rev. Lett. {\bf 95}, 170410 (2005).
\bibitem{ertmer2005} T.Sculte, S.~Drenkelforth, J.Kruse, W.Ertmer, J.~Arlt,
        K.~Sacha, J.Zakrzewski, and M.~Lewenstein,
        Phys. Rev. Lett. {\bf 95}, 170411 (2005).
\bibitem{aspect2006} D.~Cl\'ement, A.F.~Var\'on,  J.A.~Retter, L.~Sanchez-Palencia,
        A.~Aspect, and P.~Bouyer, New J. Phys. {\bf 8}, 165 (2006).
\bibitem{pitaevskii} F.~Dalfovo, S.~Giorgini, L.P.~Pitaevskii, and
         S.~Strigani, Rev. Mod. Phys. {\bf 71}, 463 (1999).
\bibitem{skipetrov2000} S.E.~Skipetrov, and 
        R.~Maynard, Phys. Rev. Lett. {\bf 85}, 736 (2000).
\bibitem{segev2007} T.~Schwartz, G.~Bartal, S.~Fishman, and M.~Segev,
        Nature {\bf 446}, 52 (2007).
\bibitem{james2005} G.~Iooss, and G.~James, Chaos {\bf 15}, 015113 (2005).
\bibitem{pikovsky1991} F.~Benvenuto, G.~Casati, A.S.Pikovsky, and D.L.Shepelyansky,
        Phys. Rev. A {\bf 44}, R3423 (1991).
\bibitem{phillips2006} C.~Ryu, M.F.~Andesen, A.~Vaziri, M.B.~d'Arcy, J.M.~Grossman,
        K.~Helmerson, and W.D.~Phillips, Phys. Rev. Lett. {\bf 96}, 160403 (2006).
\bibitem{summy2006} G.~Behinaein, V.~Ramareddy, P.~Ahmadi, and G.S.~Summy,
        Phys. Rev. Lett. {\bf 97}, 244101 (2006).
\bibitem{sadgrove2007} M.~Sadgrove, M.~Horikoshi, T.~Sekimura, and K.~Nakagawa,
        Phys. Rev. Lett. {\bf 99}, 043002 (2007).
\bibitem{Fishman82}S.~Fishman, D.~R.~Grempel, and R.~E.~Prange, 
        Phys. Rev. Lett. {\bf 49}, 509 (1982).
\bibitem{Shepelyansky-87} D.~L.~Shepelyansky, Physica D {\bf 28}, 103 (1987).
\bibitem{dls1993} D.L.~Shepelyansky, Phys. Rev. Lett. {\bf 70}, 1787 (1993).
\bibitem{kottos2004} T.~Kottos, and M.~Weiss, Phys. Rev. Lett. {\bf 93}, 190604 (2004). 
\bibitem{kramer1993} B.~Kramer, and A.~MacKinnon, Rep. Prog. Phys. {\bf 56}, 1469 (1993).
\bibitem{aspect2007} L.~Sanchez-Palencia,  D.~Cl\'ement, P.~Lugan, P.~Bouyer,
         G.V.Shlyapnikov, and A.~Aspect,  Phys. Rev. Lett. {\bf 98}, 210401 (2007).
\bibitem{shapiro2007} B.~Shapiro, Phys. Rev. Lett. {\bf 99}, 060602 (2007).
\bibitem{numrec} W.~H.~Press {\em et al.},
{\em Numerical Recipes: the Art of Scientific
                 Computing},
Cambridge University Press (1992).
\bibitem{tobepub} S.~Tietsche and A.~Pikovsky (in preparation).

\end{thebibliography}

\end{document}